\documentclass[prb,twocolumn,showpacs,preprintnumbers,amsmath,amssymb]{revtex4}
\usepackage{graphicx}
\usepackage{dcolumn}
\usepackage{bm}
\usepackage[tight]{subfigure}
\usepackage{enumerate}
\usepackage{amsmath,amssymb,amsfonts}
\usepackage{verbatim}
\usepackage{color}
\usepackage{mathrsfs}
\usepackage{setspace}
\usepackage{mathtools}
\usepackage{braket}
\usepackage{lipsum}

\begin{document}
\title{Enhanced thermoelectric efficiency of zigzag bilayer phosphorene nanoribbon; edge states engineering}
\author{Shima Sodagar}
\author{Hossein Karbaschi}
\email{h.karbaschi@gmail.com}
\author{Morteza Soltani}
\email{mo.soltani@sci.ui.ac.ir}
\author{Mohsen Amini}
\affiliation{Department of Physics, University of Isfahan, Isfahan 81746-73441, Iran} 

\begin{abstract}
We theoretically investigate the thermoelectric properties of zigzag bilayer phosphorene nanoribbons (ZBPNR). 
We first, draw an analogy between the extended Su-Schrieffer-Heeger (SSH) ladder and ZBPNR edge states and obtain their corresponding band structure and wave functions analytically. 
Then, by applying the energy filtering method, we show that the electric power and thermoelectric efficiency of the ZBPNRs can be improved remarkably in the presence of mid-gap edge states.
We also argue how to engineer the edge modes to further optimize thermoelectric power and efficiency of the system by applying periodic point potentials at the boundaries. 
\end{abstract}


\maketitle

\section{Introduction}
With the increase in world population and the advancement of technology, the need for energy resources is increasing day by day, but fossil fuel resources continue to decline. More than $60 \%$ of the input energy of industrial processes is lost as waste heat. The thermal energy is also wasted from vehicles, electronic devices like computers and mobile phones and even the human body. If this huge amount of wasted energy is converted back into usable energy, the problems of energy crisis will be solved to some extent. 
Thermoelectric power generation is considered as one of the best solutions for remarkably reducing the use of fossil fuels and preventing an energy crisis. Thermoelectric power generators (TEGs) capture thermal energy from the environment and convert it directly into electricity \cite{RoweTEbook, Goldsmid}. TEGs are reliable and maintenance-free solid-state devices that cause no noise and environmental pollution. TEGs have been studied for many years, but their use was practically very limited until the last decade. Despite the many advantages of TEG over conventional power generators, TEGs suffer from inadequate output power and low efficiency. Many studies have shown that using low-dimensional materials for thermoelectric purposes is an efficient solution to overcome the problem of low efficiency in TEGs \cite{Hicks93a, Hicks93b, Balandin2003, Murphy08, Esposito09, Leijnse10, Karbaschi2016, ZHANG2016, Hung2016, Karbaschi2020, Rezaei2021}. The Improved efficiency in low-dimensional thermoelectric modules has two main reasons. First is the existence of novel quantum effects related to the confinement of carriers. The next reason is the increase of boundaries in the system, which increases the scattering of phonons and reduces the thermal conductivity.

Two-dimensional (2D) materials refer to classes of low-dimensional crystalline solids involving a single or few layers with weak inter-layer bondings and strong intra-layer interactions in a layer. 2D materials have been widely investigated because of their special structures and unique properties which are very different from their corresponding 3D materials. This sort of materials have extensive applications due to their novel transport and optical properties. Recently, the few-layer form of black phosphorus, namely phosphorene, has been exfoliated successfully \cite{Li2014, lu2014}, and attracted huge interest due to its outstanding optical, mechanical, electronic, and thermoelectric properties. Unlike graphene, which has a zero-energy bandgap, phosphorene is a semiconductor with a large energy bandgap that can be adjusted by defects, electric field, and strain. The energy bandgap can also be modulated by the number of layers in a stack that makes bilayer phosphorene more interesting than its monolayer~\cite{TAKAO1981, Dai2014, Carvalho2016}. In addition to the high carrier mobility and large band-gap, bilayer phosphorene is easy to fabricate, which makes it more attractive to study.

The presence of edges in the ribbons considerably affects the electronic structure of phosphorene. It has been pointed out a prominent feature of phosphorene nanoribbons with zigzag edges is the presence of quasi-flat edge modes in the middle of the bandgap. Mid-gap states have distinctive properties from the bulk band and play an important role in conduction. Each phosphorene layer grants a band of two-fold degenerate edge modes.
So, two and four branches are appearing in the bandgap of energy for monolayer and bilayer zigzag phosphorene nanoribbons
correspondingly. 

One way to improve thermoelectric properties is to filter electrons at specific energies by means of a boxcar transmission function. Previous studies have shown that such a transmission function can block electrons that are responsible for the reduction of electrical output power and then increase thermoelectric efficiency \cite{Whitney14, Whitney15, Karbaschi2016}.  

The mid-gap edge states transmission function of some nanoribbons probably is the ideal boxcar function. There are usually two issues with using this transmission function. The first drawback is the small energy gap and the overlap of the edge states with the valence or conduction bands. The second issue for most cases is that the width of this transmission function is much larger than that which is suitable for use for thermoelectric applications\cite{Karbaschi2020, Rezaei2021}. 
The appropriate width for the transmission function is determined in proportion to the operating temperatures of the leads. In this study, we have proposed that the application of periodic on-site potentials at the edges of zigzag bilayer phosphorene nanoribbons can tune the width of edge states' transmission function and introduce this structure as a worthy candidate for thermoelectric purposes.

This article is organized as follows. In section \ref{sec:model}, we generally introduce the bilayer phosphorene nanoribbons crystal structure with inter-layer and intra-layer hopping parameters and introduce the TB Hamiltonian in order to study band structure. The rest of section \ref{sec:model} is devoted to discussion on zigzag bilayer phosphorene nanoribbons wave functions using the analogy between two-dimensional S-S-H ladder and bilayer phosphorene nanoribbons. In the following, we deal with the effect of all hopping parameters including intra-layer and inter-layer on wave function and dispersion of edge modes.
In section \ref{sec:thermoelectric}, methods for calculating thermoelectric properties are introduced and in section \ref{sec:results}, we presente the results of numerical study of the thermoelectric properties of zigzag bilayer phosphorene in the presence of periodic on-site potentials at both edges using the Landauer-B$\mathrm{\ddot{u}}$ttiker formalism. Lastly, we finish the article with the conclusion in section \ref{sec:Conclusion}.

\begin{figure}[]
  	\includegraphics[height=2.0\linewidth]{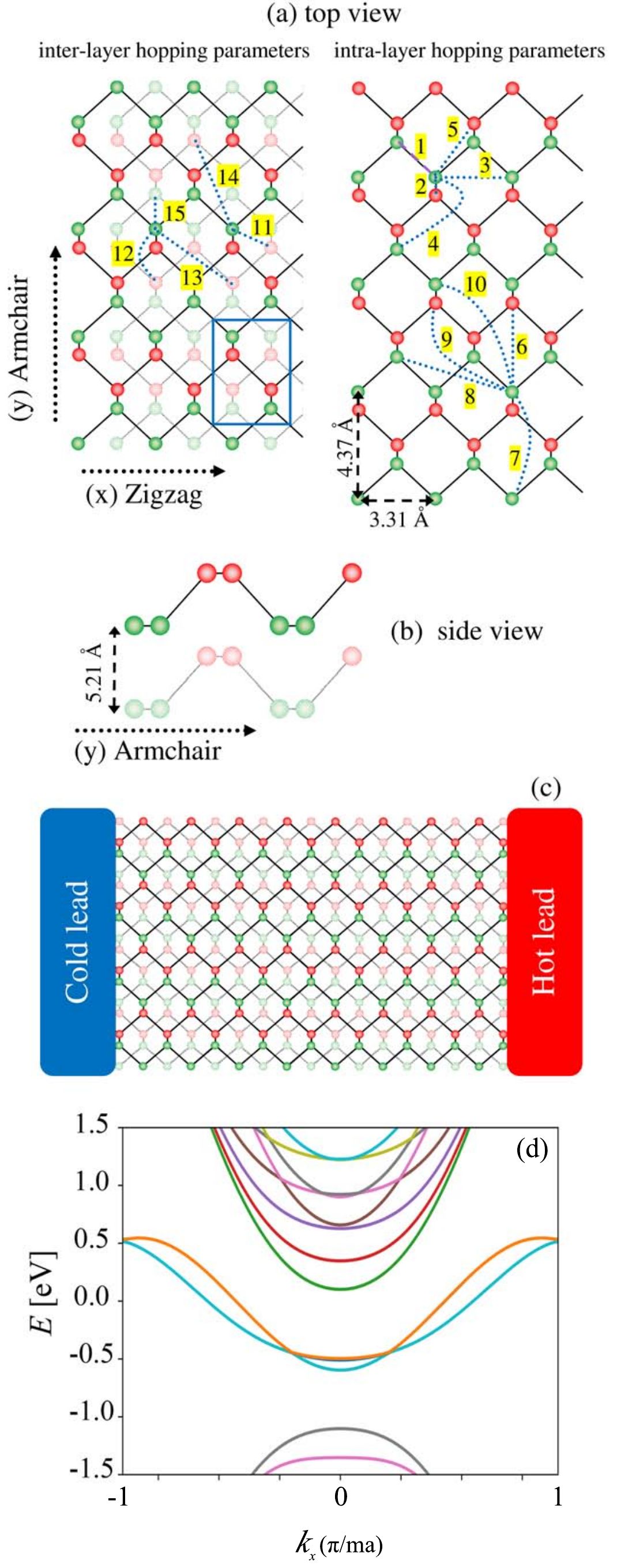}	
	\caption{\label{fig.bilayer}
	(a) five inter-layer and ten intra-layer hopping parameters in top view of bilayer phosphorene crystal structure. The blue rectangle indicates the unit cell. (b) Bilayer phosphorene crystal structure from th side view. (c) Schematic of considered system. (d) Band structure of zigzag bilayer phosphorene.
	}
\end{figure}

\section{Edge states in bilayer Phosphorene \label{sec:model}}
\subsection{Bilayer phosphorene structure}
Bilayer Phosphorene can be generally described as a two-dimensional material in which two puckered structure monolayer of Phosphorene are stacked in parallel. The layers of covalently bonded phosphorus atoms are stuck weakly to each other by Van der Waals interactions. 
Bilayer phosphorene can be considered with different stacking orders. In the A-B stacking which is considered in this study, the top layer is moved by half of a unit cell along x or y with respect to the bottom layer, figures \ref{fig.bilayer}(a) and (b).

In this study, we focus on nanoribbons with zigzag edges. A monolayer of phosphorene includes two puckered-structure sublayers. The unit cell of the phosphorene monolayer contains four atoms and for an extra layer, four additional atoms are added to the unit cell. The tight-binding Hamiltonian model is a proper technique to investigate the qualitative electronic properties of phosphorus based systems \cite{Moldovan2017,Rudenko2016}:
\begin{equation}
\begin{aligned}
H={}&\sum_{\langle i,j\rangle \langle i^\prime,j^\prime \rangle}t^{\parallel}_\mathbf{i}(\hat{c}^{\dag A_{1(2)}(B_{1(2)})}_{i,j}\hat{c}^{B_{1(2)}(A_{1(2)})}_{i^\prime,j^\prime}\\
&+\sum_{\langle i,j \rangle \langle i^\prime,j^\prime \rangle}t^{\perp}_\mathbf{i}(\hat{c}^{\dag A_1(B_2)}_{i,j}\hat{c}^{B_2(A_1)}_{i^\prime,j^\prime}\\
&+\sum_{\langle i,j \rangle}\epsilon_{i,j}(\hat{c}^{\dag A_{1(2)}(B_{1(2)})}_{i,j}\hat{c}^{A_{1(2)}(‌‌B_{1(2)})}_{i,j}),
\end{aligned}
\label{HAM}
\end{equation}
where the first and second terms describe the intra-layer and inter-layer Hamiltonian respectively.
In order to avoid confusion let us label each site by two indexes $(i, j)$  where $i$ denotes the row number along the $x$-direction and $j$ denotes the position of the site along the $y$-direction.
In Hamiltonian of Eq.~(\ref{HAM}), $t_\mathbf{i}$ describe hopping parameters between the $\langle i,j \rangle$ and $\langle i^\prime,j^\prime \rangle$ sites  which are shown in figure \ref{fig.bilayer}(a). 
$\epsilon_{i,j}$ is the on-site energy and there are set to zero for all lattice sites when no defect and external field exist in the lattice. 
The band structure of a ZBPNR can be well described by considering 15 adequate tight-binding hopping integrals (Table\ref{tab.hopping}), 5 inter-layer and 10 intra-layer \cite{Soleimanikahnoj2017}.

 In what follows we consider a ZBPNR with periodic-boundary condition in the x-direction and open boundary condition in the y-direction. Therefore, we can use the wave number in the $x$-direction, $k_x$, as a good quantum number to represent the band structure of the system.
 Figure~\ref{fig.bilayer} (d) illustrates the energy band structure of the ZBPNR with $N_w=50$ zigzag chains along the width of ribbon using the above-mentioned hopping parameters.

\begin{table}[h!]
\centering
\caption{\footnotesize 10 intra-layer ($t^{\parallel}$) and 5 inter-layer ($t^{\perp}$) hopping parameters of bilayer phosphorene, shown in figure \ref{fig.bilayer}(a).}
\label{tab.hopping}
\begin{tabular}{|c|c|c|c|c|c|c|c|c|}
\hline
$\mathbf{i}$ & $t^{\parallel}_\mathbf{i}$ &$d_\mathbf{i}(\mathrm{\AA})$&$\mathbf{i}$ & $t^{\parallel}_\mathbf{i}$ &$d_\mathbf{i}(\mathrm{\AA})$&$\mathbf{i}$ & $t^{\perp}_\mathbf{i}$ &$d_\mathbf{i}(\mathrm{\AA})$\\
\hline1&-1.486&2.22&6&0.186&4.23&11&0.524&3.60\\
2&3.729&2.24&7&-0.063&4.37&12&0.180&3.81\\
3&-0.252&3.31&8&0.101&5.18&13&-0.123&5.05\\
4&-0.071&3.34&9&-0.042&5.37&14&-0.168&5.08\\
5&0.019&3.47&10&0.073&5.49&15&0.005&5.44\\
\hline
\end{tabular}
\end{table}

\subsection{Analogy between effective SSH ladder model and bilayer phosphorene edge states}
In this section, we introduce a correspondence between  an effective SSH ladder model and ZBPNR system to understand the behavior of  edge states analytically.
 We discuss how the analogy between such systems may be used in order to find the dispersive properties of the edge bands in ZBPNRs which will be used in the following sections.
 Due to translational invariant along the $x$-direction, we can make the following Fourier transform:
\begin{equation}
\hat{c}^\dag_{(i,j)}=\sum_{k_x}e^{ik_xi}\hat{c}^\dag_{k_x,j}
\end{equation}
\begin{equation}
\hat{c}_{(i,j)}=\sum_{k_x}e^{-ik_xi}\hat{c}_{k_x,j}
\end{equation}
Since our main goal is to obtain the the edge-states behavior, we restrict our focus on the hopping parameters of the problems which do not break the particle-hole symmetry of the system. 
In this regard, the most important hopping parameters are $t^{\parallel}_1$ and $t^{\parallel}_2$ and  $t^{\perp}_{11}$ and hence, we can write our effective Hamiltonian of the ZBPNR system as :
\begin{equation}
H=H_1+H_2+H_{11},
\end{equation}
in which
\begin{equation}
H_1=\sum_{<i,j>}t^{\parallel}_1\hat{c}^{\dag A}_{(i,j+1)}\hat{c}^{B}_{(i,j)}+t^{\parallel}_1\hat{c}^{\dag A}_{(i+1,j+1)}\hat{c}^{B}_{(i,j)},
\end{equation}
\begin{equation}
H_2=\sum_{<i,j>}t^{\parallel}_2\hat{c}^{\dag A}_{(i,j)}\hat{c}^{B}_{(i,j)},
\end{equation}
and
\begin{equation}
H_{11}=\sum_{<i,j>}t^{\perp}_{11}\hat{c}^{\dag A_1}_{(i,j)}\hat{c}^{B_2}_{(i-1,j)}+t^{\perp}_{11}\hat{c}^{\dag A_2}_{(i,j)}\hat{c}^{B_1}_{(i,j)}.
\end{equation}

After performing the Fourier transformation the resulting Hamiltonian $H(k_x)=H_1(k_x)+H_2(k_x)+H_{11}(k_x)$ will be obtained as:
\begin{equation}
H_1(k_x)=\sum_{k_x}\sum_{j}\tau^{\parallel}_1(k_x)\hat{c}^{\dag A}_{(k_x,j+1)}\hat{c}^{B}_{(k_x,j)}+\tau^{\parallel}_1\hat{c}^{\dag A}_{(k_x,j+1)}\hat{c}^{B}_{(k_x,j)}e^{ik_x}),
\end{equation}
\begin{equation}
H_2(k_x)= \sum_{k_x}\sum_{j}\tau^{\parallel}_2(k_x)\hat{c}^{\dag A}_{(k_x,j)}\hat{c}^{B}_{(k_x,j)},
\end{equation}
and
\begin{equation}
H_3(k_x)=\sum_{k_x}\sum_{j}(\tau^{\perp}_{11}(k_x)\hat{c}^{\dag A_1}_{(k_x,j)}\hat{c}^{B_2}_{(k_x,j)}e^{-ik_x}+\tau^{\perp}_{11}\hat{c}^{\dag A_2}_{(k_x,j)}\hat{c}^{B_1}_{(k_x,j)}),
\end{equation}
where
\begin{equation}
\tau^{\parallel}_1(k_x)=2t^{\parallel}_1\cos(\frac{k_x}{2}),
\end{equation}
\begin{equation}
\tau^{\parallel}_2(k_x)=t^{\parallel}_2, 
\end{equation}
and
\begin{equation}
\tau^{\perp}_{11}(k_x)= t^{\perp}_{11}e^{(\frac{-ik_x}{2})}.
\end{equation}
Here we used gauge transformation$(\hat{c}^B_{k_x,j}\rightarrow \hat{c}^B_{k_x,j} e^{(\frac{-ik_x}{2})})$ and our results  indicate that new hopping parameters $\tau^{\parallel}_1(k_x)$ and $\tau^{\perp}_{11}(k_x)$ are now momentum dependent.

\begin{figure}[]
  	\includegraphics[height=1.2\linewidth]{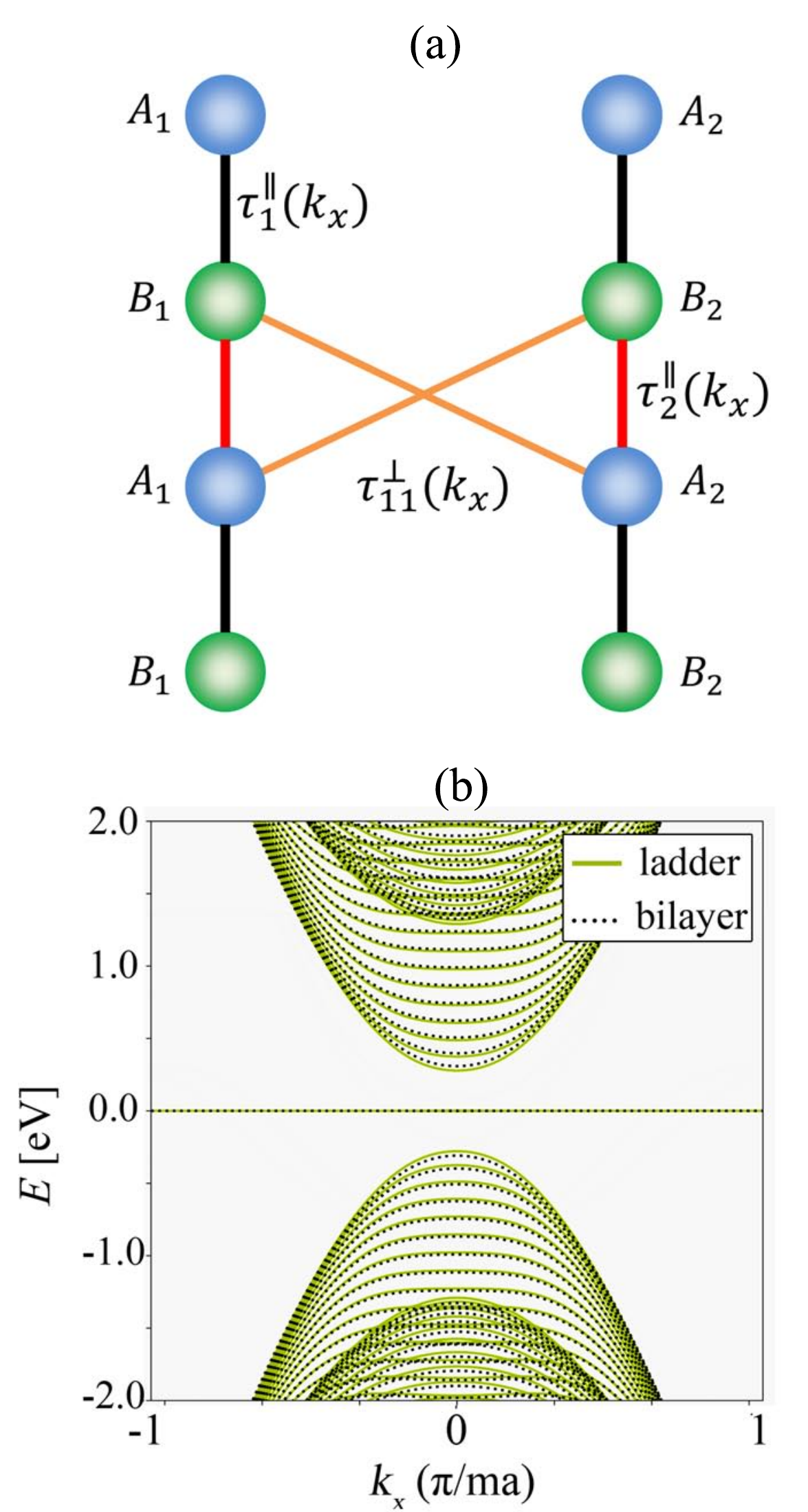}	
	\caption{\label{ssh1}
	(a) The schematic representation of the SHH ladder model in which the momentum-dependent  hopping parameters  $\tau^{\parallel}_1(k_x)$, $\tau^{\parallel}_2(k_x)$  and $\tau^{\perp}_{11}(k_x)$ are shown explicitly. (b) Comparison between band structure of the ZBPNR  and its corresponding ladder analogy with 2 intra-layer and 1 inter-layer hopping parameters.
	}
\end{figure}

The most interesting property of the resulting Hamiltonian  $H(k_x)$  is its equivalence to an effective SSH ladder with the effective hopping parameters shown in Fig~\ref{ssh1} (a). 
In order to illustrate this analogy, we have plotted the resulting band structures of both   ZBPNR system and its corresponding SSH ladder in Fig.~\ref{ssh1} (b).
As it is shown, the resulting band structure derived from the effective SSH ladder coincides with one obtained for the ZBPNR system with width $N_w=50$.
This means that our SSH ladder can be used as an analogy to represent the edge states of  ZBPNR at the middle of its energy spectrum.
Here it is appropriate to mention that we found a similar analysis to introduce an effective SSH ladder for the case of graphene nanoribbons  in Ref.~\cite{Tan2021} when we were 
writing the paper.

\subsection{Wave functions of ZBPNR edge states}
We can now proceed to obtain an analytical expression for the wave functions associated with the zero-energy state in ZBPNR. For simplicity, we generalize the notation of Refs.~\cite{Amini2019_1,Amini2019_2} for the ZBPNR to show the components of the the wave functions  in real space. As we know, each unit cell consists of two different atoms, denoted as  $A$ and $B$ which can be used to label the wave function of the atom on the outermost sites by $\psi_{n,A}$ and $\psi_{n,A}$ where $n$ shows the cell index on the edge of ribbon. We start our analysis by  writing the Schrödinger equation as:
\begin{equation}
\hat{H}\ket{\psi}=E\ket{\psi}
\end{equation}
There are $N_w$ atoms across the width of ribbon in each layer and total wave function is labeled as $\psi=\{\psi_{1,A_1},\psi_{1,A_2},\psi_{1,B_1},\psi_{1,B_2},\ldots\}$. 
We are interested in the edge states at $E=0$ which means that we need to solve the equation $\hat{H}\ket{\psi}=0$ with the following Hamiltonian: 
\begin{equation}
\hat{H}(k) =\begin{psmallmatrix} 
0 &0& \tau^{\parallel}_1(k_x)&0&0&0&\ldots \\
0 & 0&0&\tau^{\parallel}_1(k_x)&0&0&\ldots \\
(\tau^{\parallel}_1(k_x))^{\ast}&0&0&0&(\tau^{\parallel}_2(k_x))^{\ast}&(\tau^{\perp}_{11}(k_x))^{\ast}&\ldots\\
0&(\tau^{\parallel}_1(k_x))^{\ast}&0&0&(\tau^{\perp}_{11}(k_x))^{\ast}&(\tau^{\parallel}_2(k_x))^{\ast}&\ldots\\
0&0&\tau^{\parallel}_2(k_x)&\tau^{\perp}_{11}(k_x)&0&0&\ldots\\
0&0&\tau^{\perp}_{11}(k_x)&\tau^{\parallel}_2(k_x)&0&0&\ldots\\
\vdots&\vdots&\vdots&\vdots&\vdots&\vdots&
\end{psmallmatrix}
\quad
\end{equation}

We propose the following ansatz for the edge states wave function:
\begin{equation}
\begin{split}
 \ket{\psi(k_x)}_{edge}&=\sum_n(\psi_{(n,A_1)(k_x)}\ket{n,A_1}+\psi_{(n,B_1)(k_x)}\ket{n,B_1}\\
&+\psi_{(n,A_2)(k_x)}\ket{n,A_2}+\psi_{(n,B_2)(k_x)}\ket{n,B_2}),
\end{split}
\end{equation}
and try to solve the eigenvalue problem using the recursive procedure mentioned in Ref.~\cite{Amini2019_2}.
It is clear that the amplitudes of the wave functions which are localized at the ribbon's top edge are zero on even sites and similar conditions hold for the bottom edge wave functions and odd sites yielding two sets of answers.
In order to proceed further we need to consider the following two different conditions.
If the amplitudes of wave function are equal on both layers, $\psi_{n,A_1,k_x}=\psi_{n,A_2,k_x}$, the solution is symmetric and to that end:
\begin{equation}
\psi^+_{(n,A_{1},k_x)}=(\frac{-\tau^{\parallel}_1(k_x)}{\tau^{\parallel}_{2}(k_x)+(\tau^{\perp}_{11}(k_x))^{\ast}})^n\psi^+_{(0,A_{1},k_x)}
\end{equation}
By the same token, the solution is asymmetric if the amplitudes of the wave function on different layers is opposite to each other, $\psi_{n,A_1,k_x}=-\psi_{n,A_2,k_x}$, and therefore:
\begin{equation}
\psi^-_{(n,A_{1},k_x)}=(\frac{-\tau^{\parallel}_1(k_x)}{\tau^{\parallel}_{2}(k_x)-(\tau^{\perp}_{11}(k_x))^{\ast}})^n \psi^-_{(0,A_{1},k_x)}.
\end{equation}
Now, we expect the $\ket{\psi^{+(-)}{(k_x)}}_{(edge)}$ to be normalized. This implies that$\langle{\psi^{+(-)}}\ket{\psi^{+(-)}}_{edge}=1$.

\subsection{The effect of other hopping parameters on wave function and dispersion of edge-bands}
\begin{figure}[]
  	\includegraphics[height=1.2\linewidth]{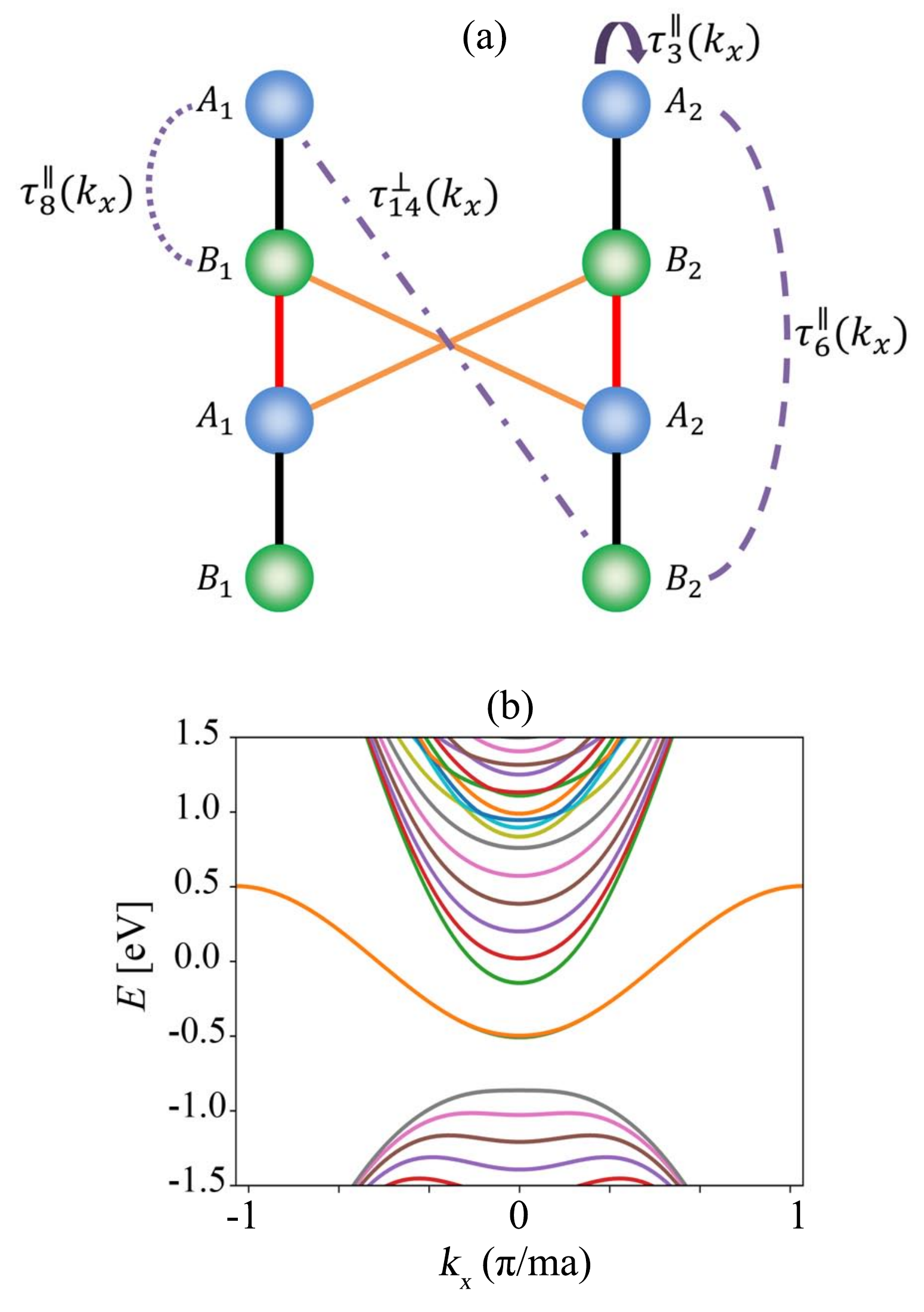}	
	\caption{\label{ssh2}
(a) The same as figure~\ref{ssh1} (a), but for the relevant new hopping parameters $\tau^{\parallel}_3(k_x)$, $\tau^{\parallel}_6(k_x)$ , $\tau^{\parallel}_8(k_x)$  and $\tau^{\perp}_{14}(k_x)$. (b) The same as figure~\ref{ssh1} (b), but in presence of hopping integral $t_3$ which is responsible for the dispersion of the edge bands.
	}
\end{figure}

The resulting edge bands obtained in the previous sub-section is not dispersive. However, in this section we consider the effect of other hopping parameters which we ignored before and discuss their effects on the dispersion of our resulting edge bands.
Therefore this section starts out with a quick look at
Table\ref{tab.hopping}, which shows that except for the main hopping parameters (represented in figure~\ref{ssh1} (a)), there are some considerable hopping parameters that may have an influence on the wave function and dispersion of  the edge band. 
Such important hopping parameters are shown in figure~\ref{ssh2} (a) by dash/dot lines.
We first, turn our attention to  the hopping integral $t^{\parallel}_6$ which has a relatively larget value among the intra-layer hopping parameters.
This hopping integral connects inequivalent atoms at the upper and lower layers and can not change the  dispersion of the edge bands. 
This is because as we have already discussed, the amplitudes of wave function is zero on even sites. However, it is interesting to calculate its effect on the analytic form of the wave function.

In order to calculate an analytic form of the wave funcion in presence of $t^{\parallel}_6$ , one needs to obtain the effective momentum dependent parameter   $(\tau^{\parallel}_6(k_x)= t^{\parallel}_6exp^{-ik_x/2})$ and then by solving the same Schrodinger equation for the edge states can obtain the  following symmetric and asymmetric solutions:
\begin{equation}
\psi^+_{(1,A_1,k_x)}=\frac{-\tau^{\parallel}_1(k_x)}{\tau^{\parallel}_2(k_x)+(\tau^{\perp}_{11}(k_x))^{\ast}}\psi^+_{(0,A_1,k_x)}
\end{equation}
\begin{equation}
\begin{split}
\psi^+_{(n,A_1,k_x)}&=\frac{(-\tau^{\parallel}_6(k_x))^{\ast}}{\tau^{\parallel}_2(k_x)+(\tau^{\perp}_{11}(k_x))^{\ast}}\psi^+_{(n-2,A_1,k_x)}\\ &+\frac{-\tau^{\parallel}_1(k_x)}{\tau^{\parallel}_2(k_x)+(\tau^{\perp}_{11}(k_x))^{\ast}}\psi^+_{(n-1,A_1,k_x)}
\end{split}
\end{equation}
\begin{equation}
\psi^-_{(1,A_1,k_x)}=\frac{-\tau^{\parallel}_1(k_x)}{\tau^{\parallel}_2(k)-(\tau^{\perp}_{11}(k_x))^{\ast}}\psi^-_{(0,A_1,k_x)}
\end{equation}
\begin{equation}
\begin{split}       
\psi^-_{(n,A_1,k_x)}&=\frac{(-\tau^{\parallel}_6(k_x))^{\ast}}{\tau^{\parallel}_2(k)-(\tau^{\perp}_{11}(k_x))^{\ast}}\psi^-_{(n-2,A_1,k_x)}\\ &+\frac{-\tau^{\parallel}_1(k_x)}{\tau^{\parallel}_2(k_x)-(\tau^{\perp}_{11}(k_x))^{\ast}}\psi^-_{(n-1,A_1,k_x)}.
\end{split}
\end{equation}

Another hopping parameters which seems to be significant is $t^{\perp}_{14}$.  
It does not eliminate the degeneracy of edge bands and the effect that the mentioned parameter has on the deformation of the wave function is exactly the same as the previous recurrence equation considering that $\tau^{\perp}_{14}(k_x)=t^{\perp}_{14}e^{(-ik_x/2)}$. $t^{\parallel}_8$ is another relevant hopping parameter that we would discuss. This is a hopping parameter that connects inequivalent sub-lattices from the upper layer of one chain to an atom in lower layer at a neighboring chain. The presence of this parameter, similar to $t^{\parallel}_6$ and $t^{\perp}_{14}$, does not remove the degeneracy of edge bands and consequently  only affects the $t^{\parallel}_1$ parameter by a small phase change.

The last hopping parameter which we consider here is  $t^{\parallel}_3$.
It is an important parameter since it is responsible for the dispersion of edge bands. 
This parameter connects the nearest neighbor sites of a pair of zigzag chains in the lower or upper layers. 
To understand its effect on the edge band dispersion, we need to perform the same analysis as before which results in the hopping form $\tau^{\parallel}_3(k_x)= 2t^{\parallel}_3\cos(k_x)$. 
On the other hand, the presence of $t^{\parallel}_3$ shifts the edge states spectrum with $2t^{\parallel}_3\cos(k_x)$ but the electron-hole symmetry around fermi level$(E=0)$ will n't be broken as well as the degeneracy of edge bands.   
The results which is plotted in figure~\ref{ssh2}(b) indicates that the flatness of the edge bands will be completely removed  by the presence of $t_3$. 

\subsection{Degeneracy removal using the perturbative analysis}

\begin{figure}[]
  	\includegraphics[height=0.55\linewidth]{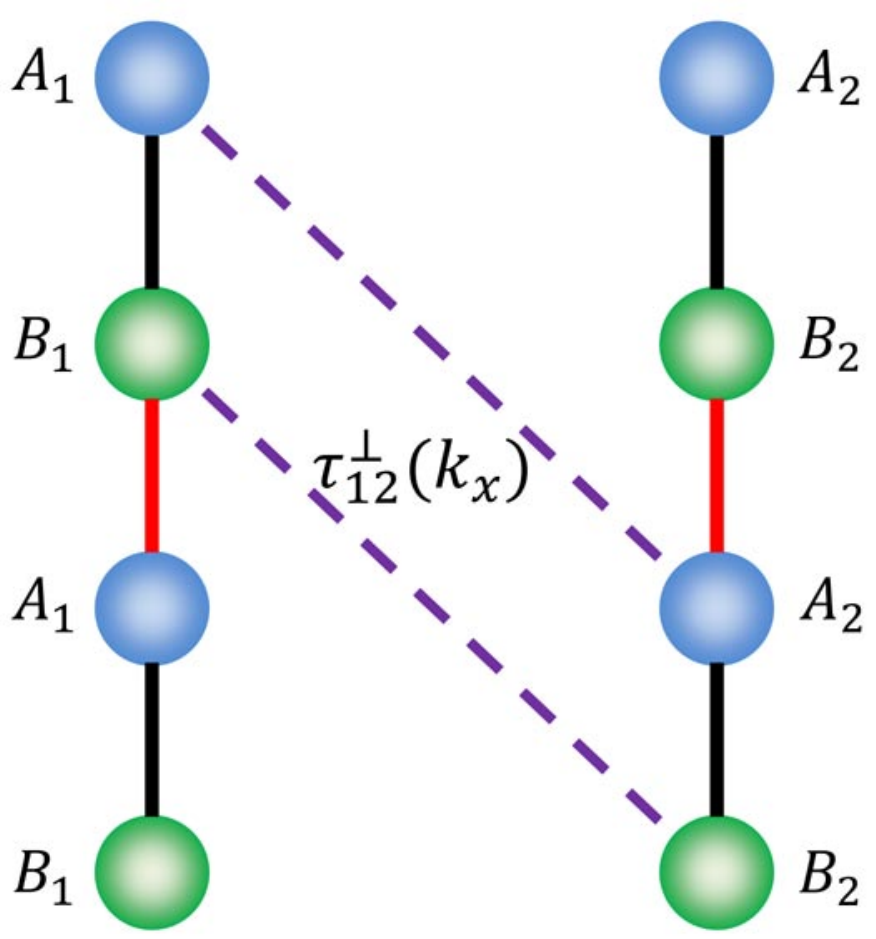}	
	\caption{Schematic representation of hopping parameter $t^{\perp}_{12}$ in the ladder analogy of the ZBPNRs.}
\label{ft12}
\end{figure}

In the remainder of this section we will present the derivation of removing the degeneracy of the  edge bands analytically via perturbation theory.
Our investigation indicates that using  the above-mentioned wave functions in the standard first-order perturbation theory and taking the expectation value of the most relevant parameters like $t^{\perp}_{12}$  and $t^{\perp}_{13}$ , we can estimate the dispersion energy of the edge bands. In order to understand its ladder analogy we have presented the parameter $t^{\perp}_{12}$ in figure~\ref{ft12}.

Each layer of zigzag phosphorene nanoribbon contributes a band of two-fold degenerate mid-gap modes, making four mid-gap modes for  ZBPNRs. In what follows we employ the degenerate perturbation theory to obtain the dispersion of edge bands for ZBPNRs.
 We consider the Hamiltonian $H$ in such a way that it can be split into two parts, namely $\hat{H}=\hat{H}_0+\hat{H^{\prime}}$, where $\hat{H^{\prime}}$ is known as the perturbation. Thus, we can write
\begin{eqnarray}
\hat{H^{\prime}}=\begin{pmatrix} 
E_{11}&E_{12}\\
E_{21}&E_{22}
\end{pmatrix}
\end{eqnarray}
and the perturbative Hamiltonian matrix elements are determined by:
\begin{eqnarray}
\begin{cases}
E_{11}=\bra{\psi^+}\hat{H^{\prime}}\ket{\psi^+}\\
E_{12}=\bra{\psi^+}\hat{H^{\prime}}\ket{\psi^-}\\
E_{21}=\bra{\psi^-}\hat{H^{\prime}}\ket{\psi^+}\\
E_{22}=\bra{\psi^-}\hat{H^{\prime}}\ket{\psi^-}.
\end{cases}
\end{eqnarray}

Then, the eigenvalues in the presence of the hopping $t^{\perp}_{12}$ and $t^{\perp}_{13}$ will be given by the following equation:

\begin{eqnarray}
\begin{vmatrix} 
E-\bra {\psi^+} \hat{H^{\prime}}\ket{\psi^+}&
 \bra {\psi^+} \hat{H^{\prime}} \ket{\psi^-}\\
\bra {\psi^-}  \hat{H^{\prime}} \ket{\psi^+}&
E-\bra {\psi^-}  \hat{H^{\prime}} \ket{\psi^-}
\end{vmatrix}=0 \quad
\end{eqnarray}

Applying the same procedure we used in the previous section and considering the gauge transformation, we try to write the perturbative Hamiltonian as:
\begin{equation}
\begin{aligned}
\hat{H^{\prime}}_{12}(k_x)={}& \sum_{k_x}\sum_{j} (t^{\perp}_{12})\{\hat{c}^{\dag A_1}_{(k_x,j+1)}\hat{c}^{A_2}_{(k_x,j+2)}\\
&+\hat{c}^{\dag A_2}_{(k_x,j)}\hat{c}^{A_1}_{(k_x,j+1)}+\hat{c}^{\dag B_1}_{(k_x,j)}\hat{c}^{B_2}_{(k_x,j+1)}e^{(i\frac{k_x}{2})}\\
&+\hat{c}^{\dag B_2}_{(k_x,j-1)}\hat{c}^{B_1}_{(k_x,j)}e^{(-i\frac{k_x}{2})}\}
\end{aligned}
\end{equation}
\begin{equation}
\begin{aligned}
\hat{H^{\prime}}_{13}(k_x)={}& \sum_{k_x}\sum_{j} (t^{\perp}_{13})2\cos(k)\{\hat{c}^{\dag A_2}_{(k_x,j)}\hat{c}^{A_1}_{(k_x,j+1)}\\
&+\hat{c}^{\dag A_1}_{(k_x,j+1)}\hat{c}^{A_2}_{(k_x,j)}+\hat{c}^{\dag B_2}_{(k_x,j-1)}\hat{c}^{B_1}_{(k_x,j-2)}\\
&+\hat{c}^{\dag B_1}_{(k_x,j-2)}\hat{c}^{B_2}_{(k_x,j-1)}\}
\end{aligned}
\end{equation}
This equation can be solved and the resulting eigen energies are the following:
\begin{equation}
E^{(1)}_{12}=t^{\perp}_{12}(\sum_j \psi^\ast_{(j+1,A_1)}\psi_{(j+2,A_2)})+t^{\perp}_{12}(\sum_j\psi^\ast_{(j,A_2)}\psi_{(j+1,A_1)})
\end{equation}
and so considering the $\frac{-\tau^{\parallel}_{1}(k_x)}{\tau^{\parallel}_{2}(k_x)+(\tau^{\perp}_{11}(k_x))^{\ast}}=x$ and $\frac{-\tau^{\parallel}_{1}(k_x)}{\tau^{\parallel}_{2}(k_x)-(\tau^{\perp}_{11}(k_x))^{\ast}}=y$, matrix elements of perturbation hamiltonian consist of:
\begin{equation}
E_{11}=t^{\perp}_{12}(x)(\frac{1+x^2}{1-x^2})
\end{equation}
\begin{equation}
E_{22}=t^{\perp}_{12}(y)(\frac{1+y^2}{1-y^2})
\end{equation}
\begin{equation}
E_{12}=t^{\perp}_{12}(\frac{y}{(1-x^{\ast})(1-y)})(x^{\ast}y+1)
\end{equation}
and about another hopping parameter:
\begin{equation}
\begin{split}
E^{(1)}_{13}&=t^{\perp}_{13}(2\cos(k_x))(\sum_j \psi^\ast_{(j,A_2)}\psi_{(j+1,A_1)})\\
&+t^{\perp}_{13}(2\cos(k_x))(\sum_j\psi^\ast_{(j+1,A_1)}\psi_{(j,A_2)})
\end{split}
\end{equation}
\begin{equation}
E_{11}=2t^{\perp}_{13}\cos(k_x)(\frac{x+(x^\ast)}{1-x^2})
\end{equation}
\begin{equation}
E_{22}=2t^{\perp}_{13}\cos(k_x)(\frac{y+(y^\ast)}{1-y^2})
\end{equation}
\begin{equation}
E_{12}=2t^{\perp}_{13}\cos(k_x)(\frac{1}{(1-x^{\ast})(1-y)})(x^{\ast}+y)
\end{equation}
This result is consistent  with  our numerical calculations and our conclusion is that hopping parameters $t^{\perp}_{12}$ and $t^{\perp}_{13}$ are the responsible parameters for the removal of the degeneracy between energy bands associated with each layer.

\section{Thermoelectric properties} \label{sec:thermoelectric}

In the thermoelectric power generation process, electrical power (output) is generated in cost of losing heat from hot lead (input). Therefore, it is desirable to produce the maximum amount of electrical power with the least heat flow. The efficiency of TEGs is then defined as:

\begin{equation}\label{eq:cop}
	\eta^{TEG}(\%) = \frac{P^{el}}{Q_{hot}}=\frac{VI}{Q_{hot}} \times100 \%,
\end{equation}

where $I$ and $Q_{hot}$ are electric and heat currents and can be obtained using the Landauer-B$\mathrm{\ddot{u}}$ttiker formalism and $V$ denotes the bias voltage. The electric current in the ballistic transport regime can be obtained from the following equation:
\begin{equation}\label{eq:current}
	I = \frac{2 q_e}{h} \int dE \; T_{HC}(E) \left( f_H-f_C\right),
\end{equation}

where $q_e$, $h$ and, $T_{HC}(E)$ denote the elementary charge, the Planck constant and, total transmission function, respectively.  Also,  $f_{H}$ and $f_{C}$ are the Fermi-Dirac distribution functions at hot and cold leads.

The heat current has two electron and lattice parts. In order to calculate the electron part of the heat current, $e$ should be replaced with the energy which each electron carries. 

\begin{equation}\label{eq:heatcurrent}
	Q_{hot} = \frac{2}{h} \int dE \; T_{LR}(E) (E - \mu_h)  \left( f_L-f_R\right).
\end{equation}

Many studies have shown that the lattice part of thermal conductivity of low-diemensional materials is much less than their corresponding bulk structures which is mainly due to increased scattering of phonons at the boundaries. Since the reduction of thermal conductivity leads to an increase in thermoelectric efficiency, this is the basis for the idea of using low-dimensional materials for thermoelectric applications. 

To further increase the efficiency, several methods have been proposed that increase the phonon scattering while having no significant destructive effects on the electronic properties. The use of superlattices \cite{Karbaschi2016} and heterojunctions \cite{Wu2020,Chen2020_2} and introducing strain \cite{Chen2020_1} and defects \cite{Rezaei2021} are the most relevant approaches to thermal conductivity reduction.

Optimizing the electronic properties of the system is another approach to enhance electric power and thermoelectric efficiency.
As mentioned earlier, the energy filtration method by means of a boxcar transmission function leads to optimal output power and efficiency \cite{Whitney14, Whitney15, Karbaschi2016}. A boxcar transmission function can eliminate the destructive current of electrons that flow from cold to hot lead and cause the enhancement of the output power.

We select zigzag bilayer phosphorene nanoribbons for this study for two reasons. The first is that among two-dimensional structures it has a small thermal conductivity, and the second is the presence of mid-gap edge modes , which provide a desired boxcar transmission function. The width of edge states transmission function in pristine bilayer zigzag phosphorene nanoribbons is very large and not suitable for room temperature thermoelectric applications. To regulate the width of this function, on-site potentials have been applied periodically at the two edges of the nanoribbon. The existence of on-site potentials also increases the phonon scattering and significantly reduces the lattice thermal conductivity of nanoribbon. In this study and for the cases of dense on-site potentials, we neglected the lattice part of thermal conductivity.

The transmission function can be calculated using the Landauer-B$\mathrm{\ddot{u}}$ttiker formalism:
\begin{equation}
T_{LR}(\epsilon)=Tr(\Gamma_{L}G^{ret}\Gamma_{R}{G^{ret}}^\dag).
\end{equation}

Here, $\Gamma_{L(R)}(\epsilon)=i[\Sigma_{L(R)}(\epsilon)-\Sigma_{L(R)}^\dagger(\epsilon)]$ is the coupling between device and left(right) lead and $G^{ret}$ is the retarded Green’s functions:

\begin{equation}
G^{ret}(\epsilon)=\frac{1}{\epsilon-H_{center}-\Sigma_L{(\epsilon)}-\Sigma_R{(\epsilon)}}
\end{equation}

where, $H_{center}$ is the device Hamiltonian and $\Sigma_{R}$ ad $\Sigma_{L}$ are the coupling to the leads retarded self energies
\begin{equation}
\Sigma_{L}(\epsilon)=\tau_{LD}^\dagger~G^{ret}_{L}~\tau_{LD},\\
\Sigma_{R}(\epsilon)=\tau_{DR}~G^{ret}_{R}~\tau_{DR}^\dagger,
\end{equation}
in which $\tau_{DR}$ and $\tau_{LD}$ are the coupling Matrices between device and right and left leads. Also, $G^{ret}_{R(L)}=[\epsilon-H_{L(R)}]^{-1}$ is the right(left) lead retarded Green's function and, $H_{L}$ and $H_{R}$ are the left and right leads Hamiltonian.

The second law of thermodynamics sets a maximum to the heat engine's efficiency operating at two temperatures. This ideal limit which is called Carnot efficiency, depends on the temperatures of the cold and hot leads and is given by:

\begin{equation}\label{eq:Carnot}
\eta_{Carnot}=1-\left(\frac{T_c}{T_h} \right)
\end{equation} 

\begin{figure*}[]
  	\includegraphics[height=0.65\linewidth]{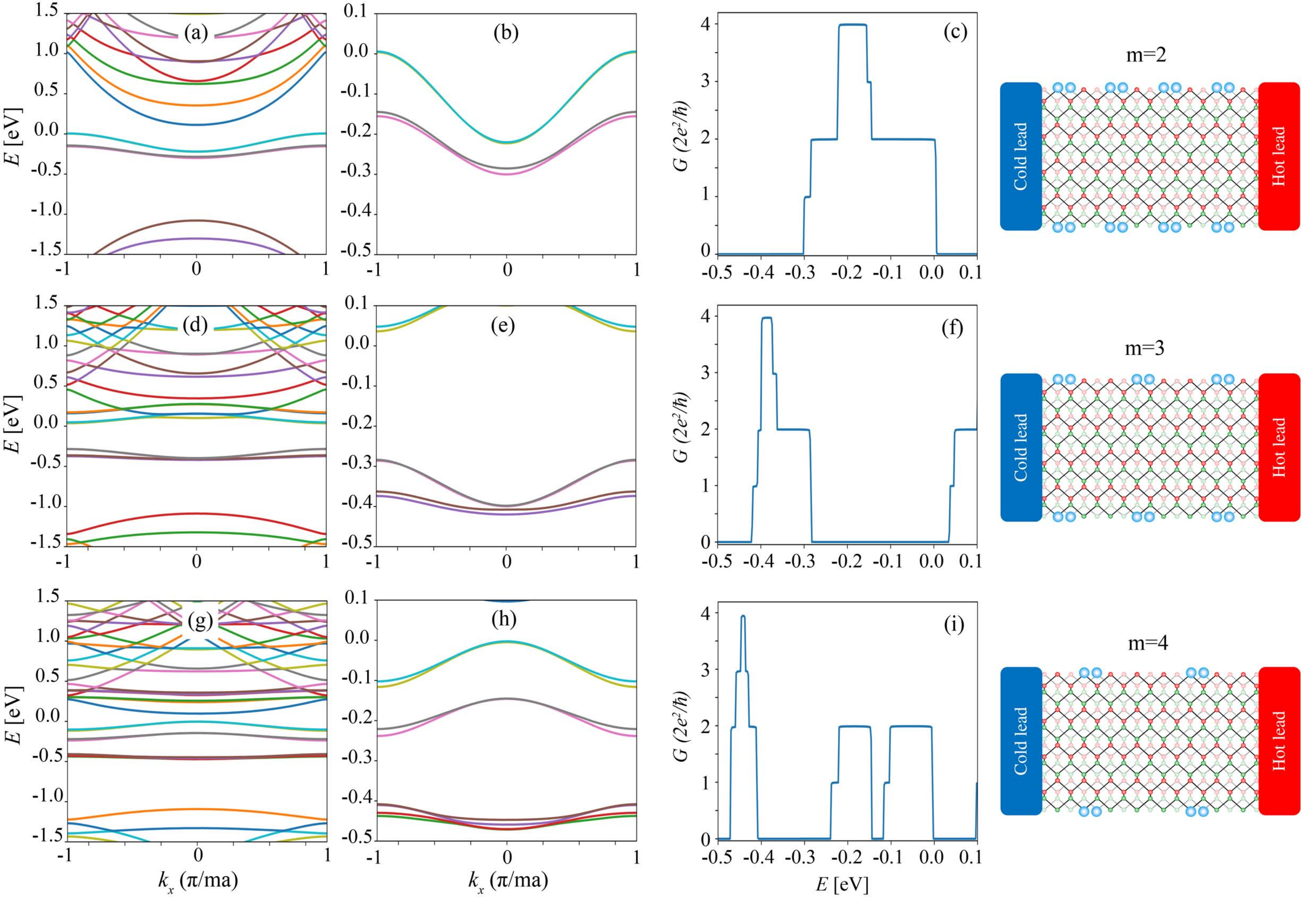}	
	\caption{\label{fig:bands}
(a), (d), and (g) band structure plotted over a wide range of energies for the cases of m=2, 3, and 4, respectively. (b), (e), and (h) show the band structure and (c) , (f) and (i) demonstrate the  transmission function close to the edge states corresponding to (a), (d), and (g), respectively. The blue circles indicate the position of on-site potentials for each case.
	}
\end{figure*}

This ideal efficiency is obtained when electrons of only one certain energy are transported, which means $T=\delta(E-E_0)$. Although very narrow transmission functions lead to high efficiencies, they provide negligible electrical power output. The use of the boxcar transmission function is the solution to optimize both electric power output and efficiency. 

Such a boxcar transmission function blocks all electrons except low-energy electrons close to Fermi energy, and since electron-phonon scattering is significant for high-energy electrons, we omit this electron-phonon interaction in this study. Also, due to the presence of many impurities in the system, the phonon thermal conductivity will be significantly reduced. In this study, we have omitted the heat flow due to phonon thermal conductivity, but it can later be added to the calculations as a disturbance

\begin{figure}[]
  	\includegraphics[height=1.5\linewidth]{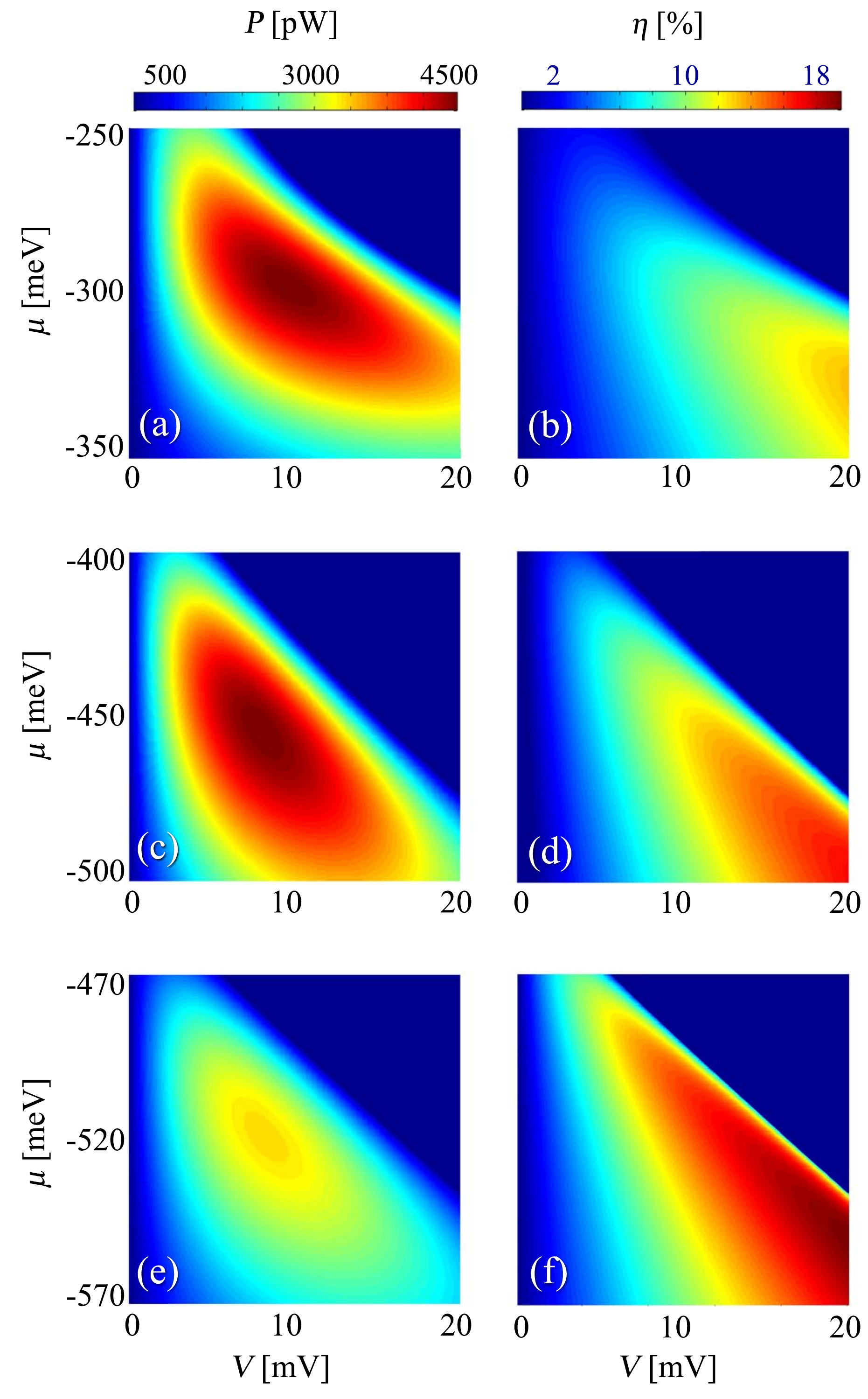}	
	\caption{\label{fig:Pandeff}
	(a), (c), and (e) electrical power output and (b), (d), and (f) the thermoelectric efficiency as a function of $V$ and $\mu$ for the cases of m=2, 3, and 4, respectively.
	}
\end{figure}

\section{Results \label{sec:results}}
In the nonlinear calculations, the hot and cold lead temperatures are assumed to be $T_h = 400~\mathrm{K}$ and $T_c = 300~\mathrm{K}$, so the ideal Carnot efficiency at these temperatures is $25 \%$. 

As shown in figure~\ref{fig:bands}, periodic on-site potentials with three periodicities are applied to both edges of the zigzag bilayer phosphorene nanoribbons. In the first case labeled with m = 2, the on-site potentials are applied to two adjacent atoms, one atom from the top layer and another atom from the bottom layer of both edges of ZBPNRs while the next two atoms have no on-site potentials. In the second and third cases with m=3 and m=4, after two atoms with on-site potentials, the next 4 and 6 atoms remain intact, respectively. 

Figures \ref{fig:bands}(a), \ref{fig:bands}(d), and \ref{fig:bands}(g) show the energy bandstructure of ZBPNRs for the cases of m=2, 3 and 4, respectively. It can be seen that there are a number of energy bands corresponding to the edge states in the energy bandgap. 

Figures \ref{fig:bands}(b), \ref{fig:bands}(e), and~\ref{fig:bands}(h) illustrate a zoom of energy bands and figures~\ref{fig:bands}(c), \ref{fig:bands}(f), and \ref{fig:bands}(i) show their corresponding transmission function close to the edge states.

We see that changing the spacing between the on-site potentials results in a shift in the position of the transmission peaks. At larger distances between the on-site potentials, the transmission peaks come closer to each other. This is similar to the situation that occurs in continuous systems such as rectangular potential barriers and makes sense because changing the distance between the on-site potentials leads to a change in resonance between their wave function. So, more transmission peaks can be observed in the bandgap for cases with larger spacing between on-site potentials. Also, increasing the spacing between the on-site potentials reduces the energy bands dispersion and ultimately leads to narrower boxcar transmission functions. Narrowing of the transmission windows by increasing the spacing between on-site potentials is caused by decreasing the overlap of their localized states.

Figures \ref{fig:Pandeff}(a), \ref{fig:Pandeff}(c), and \ref{fig:Pandeff}(e) show the electric power and figures \ref{fig:Pandeff}(b), \ref{fig:Pandeff}(d), and \ref{fig:Pandeff}(f) demonstrate the thermoelectric efficiency, using the transmission function of figures \ref{fig:bands}(c), \ref{fig:bands}(f), and \ref{fig:bands}(j), plotted on color scale as functions of bias voltage $V$ and the average chemical potential $\mu$. We can attain optimal electric power and thermoelectric efficiency by properly regulating bias voltage and average chemical potential. 
As indicated in figure \ref{fig:Pandeff}, the maximum thermoelectric efficiency is larger for cases with larger spacing between on-site potentials which have a narrower transmission functions. These results show that the output power at the maximum efficiency point is much smaller than the maximum power, so the maximum power point is probably a better choice than the maximum efficiency point. 
The highest maximum electric power is obtained for the case of $m=2$, which has the widest transmission function, $P^{m=2}_{max}=4550~\mathrm{pW}$ and for the other cases $P^{m=3}_{max}=4450~\mathrm{pW}$ and $P^{m=4}_{max}=2930~\mathrm{pW}$.
In contrast to maximum electric power, the highest efficiency at maximum electric power is is achieved for the case of m=4 which has the narrowest transmission function, $\eta^{m=4}_{maxP}=12.7\%$ and for the other cases $\eta^{m=3}_{maxP}=11.0\%$ and $\eta^{m=2}_{maxP}=8.6\%$.
Considering the values obtained for maximum electric power and its corresponding efficiency, it seems that the optimal electric power and thermoelectric efficiency for the considered temperatures are achieved for the case of m=3, whose maximum electric power is very close to the case m=2, but with higher efficiency.
\\

\section{Conclusion \label{sec:Conclusion}}
In conclusion, we have investigated the thermoelectric power generation properties of zigzag bilayer phosphorene nanoribbons in the presence of periodic on-site potentials at the edges and have calculated the generated electric power and thermoelectric efficiency. Since the width of the edge state transmission function in pristine ZBPNRs is very large, we don't expect good thermoelectric efficiency at room temperature for this structure. By periodically applying on-site potentials, we have obtained narrow transmission functions that are suitable for room temperature thermoelectric applications. Our results show that excellent electric power can be achieved at high efficiencies by edge states engineering.

\end{document}